\documentstyle[12pt]{article}

\begin{document}
\thispagestyle{empty}
{\baselineskip0pt
\leftline{\large\baselineskip16pt\sl\vbox to0pt{\hbox{DAMTP} 
               \hbox{University of Cambridge}\vss}}
\rightline{\large\baselineskip16pt\rm\vbox to20pt{
               \hbox{DAMTP-1998-142}
               \hbox{KUNS-1539}
               \hbox{UTAP-305}
               \hbox{RESCEU-54/98}
               \hbox{\today}
\vss}}%
}
\vskip20mm

\begin{center}
{\large\bf Back Reaction to the Spectrum of Magnetic Field in the
Kinetic Dynamo Theory \\
--- Modified Kulsrud and Anderson Equation ---}
\end{center}

\begin{center}
{\large Tetsuya Shiromizu
\footnote{JSPS Postdoctal Fellowship for Research Abroad}} \\
\vskip 3mm
\sl{DAMTP, University of Cambridge \\ 
Silver Street, Cambridge CB3 9EW, UK \\
\vskip 5mm
Department of Physics, The University of Tokyo, Tokyo 113-0033, 
Japan \\
and \\
Research Center for the Early Universe(RESCEU), \\ 
The University of Tokyo, Tokyo 113-0033, Japan
}
\end{center}
\begin{center}
{\large Ryoichi Nishi}
\vskip 3mm
\sl{Department of Physics, Kyoto University, Kyoto 606-8502, Japan}
\end{center}
\vskip 5mm
\begin{center}
\it{to be published in Prog. Theor. Phys.}
\end{center}

\begin{abstract}
We take account of the lowest order back reaction on the 
fluid and modify the Kulsrud and Anderson equation 
$\partial_t{\cal E}_M= 2 \gamma {\cal E}_M$ obtained 
in the kinetic dynamo theory, where ${\cal E}_M$ is 
the energy density of the magnetic field. Furthermore, we apply our 
present result to some astrophysical stages where the magnetic field 
is expected to be amplified by the dynamo mechanism.  
\end{abstract}

\section{Introduction}


The magnetic fields have been observed in various astrophysical scales
\cite{Obs}. The origin is one of the important problems in cosmology
\cite{PMF}. 
Although an attractive mechanism in protogalaxy was proposed by 
Kulsrud et al\cite{Cen} for the galactic magnetic field, it cannot 
explain how the magnetic field is made in intergalactic and
intercluster regions\cite{PMF}. Thus, it is worth investigating the 
generation and the evolution of the primordial magnetic field. 
Since the strength of these fields strength is too small we expect 
that the amplification of these fields 
occurs due to the dynamo mechanism. It is well known that 
the mean magnetic field can be amplified enough to explain 
the present observation in the kinetic dynamo theory\cite{Dynamo}. 
However, as Kulsrud and Anderson showed\cite{KA}, the growth rate of 
the fluctuation around the mean magnetic fields is much larger than that of 
the mean field in interstellar mediums. This means that 
the kinetic dynamo theory breaks down. Hence, one 
must investigate the effect of the back-reaction to the kinetic
theory. 
So far the back-reaction on the mean field has been considered\cite{MFT}.  
Setting apart the problem of the kinetic dynamo theory in interstellar
mediums, it is obvious that 
the kinetic theory cannot hold near the equipartition state in
general. 

In this paper, we consider the back-reaction on the fluctuation 
and derive the evolutional equation of energy of the magnetic 
field, that is, modified Kulsrud and Anderson equation. 
Then we apply it to some examples. 

The rest of the paper is organized as follows. In Sec. II, we 
derive the modified Kulsrud and Anderson equation with
the lowest order back-reaction under a phenomenological 
assumption. In Sec. III and IV, we give applications 
and remarks, respectively. 

\section{Modified Kulsrud and Anderson Equation}

In the Fourier space the basic equations of the 
incompressible MHD are 
%
\begin{eqnarray}
\partial_t v_i({\bf k},t)=-i P_{ijk}({\bf k})
\int \frac{d^3q}{(2 \pi)^3}\Bigl[v_j({\bf k}-{\bf q},t)v_k({\bf q},t)
-b_j({\bf k}-{\bf q},t)b_k({\bf q},t) \Bigr]
\end{eqnarray}
%
and
%
\begin{eqnarray}
\partial_t b_i({\bf k},t)=i k_j \int \frac{d^3q}{(2\pi)^3}
\Bigl[v_i({\bf k}-{\bf q},t)b_j({\bf q},t)-v_j({\bf k}-{\bf q},t)
b_i({\bf q},t)\Bigr],
\end{eqnarray}
%
where $b_i({\bf k},t):=B_i({\bf k},t)/{\sqrt {4\pi \rho}}$, 
$\rho$ is the energy density of the fluid and 
$P_{ijk}({\bf k})=k_jP_{ik}({\bf k})
=k_j(\delta_{ik}-k_ik_k/|{\bf k}^2|)$. For simplicity, we neglected 
the diffusion terms in the above equations.  
This simplification can be justified by the fact that almost of 
astrophysical systems have high magnetic Reynolds number.

Following Kulsrud and Anderson\cite{KA} and considering 
a small time step as the parameter of expansion, we evaluate the 
time evolution of the magnetic field by iterations;
%
\begin{eqnarray}
b_i({\bf k},t)& = & b_i({\bf k},0)+b_i^{(1)}({\bf k},t)
+b_i^{(2)}({\bf k},t)+\cdots \nonumber \\
& = & b_i^{(0)}({\bf k})+b_i^{(1)}({\bf k},t)
+b_i^{(2)}({\bf k},t)+\cdots,
\end{eqnarray}
%
where $b_i^{(0)}({\bf k})$ is the initial field. For the fluid 
velocity, we take account of the back-reaction from the magnetic 
field (Lorentz force) as follows;
%
\begin{eqnarray}
v_i({\bf k},t)=v_i^{(1)}({\bf k},t)+\delta v_i({\bf k},t),
\end{eqnarray}
%
where $v_i^{(1)}({\bf k},t)$ is statistically homogeneous and
isotropic component and satisfy 
%
\begin{eqnarray}
\langle v_i^{(1)*}({\bf k},t') v_j^{(1)}({\bf q},t) \rangle 
& = & (2\pi)^3\Bigl[J_1(k)P_{ij}({\bf k})+iJ_2(k)\epsilon_{ikj}k_k\Bigr] 
\delta^3({\bf k} -{\bf q}) \delta (t-t') \nonumber \\
& = & (2 \pi)^3V_{ij}({\bf k}) \delta^3 ( {\bf k}-{\bf q})\delta (t-t').
\end{eqnarray}
%
This statics holds in the region where is far from the boundary. 
$J_1(k)$ and $J_2(k)$ denote the velocity dispersion and 
the mean helicity of the fluid,
%
$$
\langle {\bf v}^{(1)}({\bf x},t) \cdot  {\bf v}^{(1)}({\bf x},t) \rangle =
2 \int \frac{d^3k}{(2\pi)^3}J_1(k)\delta (0)
$$
%
and
%
$$
\langle {\bf v}^{(1)}({\bf x},t) \cdot \nabla \times {\bf v}^{(1)}({\bf 
x},t) \rangle =-2 \int \frac{d^3k}{(2\pi)^3}k^2J_2(k)\delta (0).
$$
%
The second term in the right-hand side of the eq.(4), $\delta v_i({\bf
 k},t)$, is determined by eq. (1) and this corresponds to 
the back-reaction term from the magnetic field. 

In each orders, the MHD equation becomes 
%
\begin{eqnarray}
\partial_t b_i^{(1)}({\bf k},t)= 2ik_j 
\int \frac{d^3q}{(2\pi)^3}v_{[i}({\bf k}-{\bf q},t)b^{(0)}_{j]}
({\bf q})
\end{eqnarray}
%
%
\begin{eqnarray}
\partial_t b_i^{(2)}({\bf k},t)= 2ik_j 
\int \frac{d^3q}{(2\pi)^3}v_{[i}({\bf k}-{\bf q},t)b^{(1)}_{j]}
({\bf q},t)
\end{eqnarray}
%
and
%
\begin{eqnarray}
\partial_t v_i ({\bf k},t) = 2i P_{ijk}({\bf k}) \int
\frac{d^3q}{(2\pi)^3} 
b_{(j}^{(0)}({\bf k}-{\bf q})b^{(1)}_{k)} ({\bf q},t). 
\end{eqnarray}
%
The last equation contains an effect of the lowest order back-reaction 
on the fluid and 
it gives an explicit expression of $\delta v_i({\bf k},t)$
%
\begin{eqnarray}
\delta v_i ({\bf k},t) \simeq 2i P_{ijk}({\bf k}) \int^t_0dt' \int
\frac{d^3q}{(2\pi)^3} 
b_{(j}^{(0)}({\bf k}-{\bf q})b^{(1)}_{k)} ({\bf q},t'). 
\end{eqnarray}
%
From the eqs. (6) $\sim$ (9), the time derivative of the energy becomes
%
\begin{eqnarray}
\partial_t \langle |{\bf b}({\bf k},t)|^2\rangle
& = & \langle b_i^{(1)*}({\bf k},t){\dot b}_i^{(1)}({\bf k},t) \rangle 
+b_i^{(0)*}\langle {\dot b}_i^{(2)}({\bf k},t) \rangle +{\rm c.c.}
\nonumber \\
& = & 4\int^t_0 dt' \int \frac{d^3qd^3p}{(2\pi)^6}k_j k_k 
b_{[j}^{(0)*}({\bf q}) 
\langle v_{i]}^*({\bf k}-{\bf q},t')v_{[i}({\bf k}-{\bf p}, t)\rangle 
b_{k]}^{(0)}({\bf p}) \nonumber \\
& & -4 \int^t_0 dt' \int \frac{d^3qd^3p}{(2\pi)^6}k_j q_\ell 
b^{(0)*}_i({\bf k})\langle v_{[i}^*({\bf q}-{\bf k},t)
v_{[j]}({\bf q}- {\bf p},t') \rangle b_{\ell ]}^{(0)}({\bf p}) \nonumber \\
& & ~~~~~~+{\rm c.c.}, 
\end{eqnarray}
%
where
%
\begin{eqnarray}
\langle v_i^*({\bf k},t')v_j({\bf q},t)\rangle 
& \simeq & \langle v_i^{(1)*}({\bf k},t')v_j^{(1)}({\bf q},t)\rangle 
\nonumber \\
& & +\langle v_i^{(1)*}({\bf k},t')\delta v_j ({\bf q},t)\rangle
+\langle \delta v_i^{*}({\bf k},t') v_j^{(1)}({\bf q},t)\rangle
\nonumber \\
& = & (2\pi)^3V_{ij}({\bf k})\delta^3 ({\bf k}-{\bf q}) \delta (t-t') 
\nonumber \\
& & -4 \int^t_0dt'' \int\frac{d^3 p}{(2\pi)^3}P_{jk \ell}({\bf q})
p_m b^{(0)}_{[m}({\bf p}-{\bf k})V_{i( \ell ]}
({\bf k}) b^{(0)}_{k)}({\bf q}-{\bf p}) \nonumber \\
& & -4 \int^{t'}_0dt'' \int \frac{d^3 p}{(2\pi)^3}P_{i k \ell}
({\bf k})p_m
b^{(0)*}_{[m}({\bf p}-{\bf q})V_{(\ell ] |j|}({\bf q})
b^{(0)*}_{k)}({\bf k}-{\bf p}) \nonumber \\
& =: & (2\pi)^3V_{ij}({\bf k})\delta^3({\bf k}-{\bf q})\delta (t-t')
+\delta \langle v_i^*({\bf k},t')v_j({\bf q},t)\rangle. 
\end{eqnarray}
%
The above eq. (10) with (11) is the formal equation with the 
effect of the back-reaction. 

Let us consider the simple example with the following 
initial condition
%
\begin{eqnarray}
b^{(0)}_i({\bf x})=b_0 \delta_{i z}~~~{\rm or}~~~
b^{(0)}_i({\bf k})=b_0 (2\pi)^3\delta^3({\bf k})\delta_{i z}.
\end{eqnarray}
%
This condition holds approximately as long as the spatial scale of the
magnetic field is much larger than the typical scale of eddies. 
In this case, the eq. (10) becomes
%
\begin{eqnarray}
\partial_t \langle |{\bf b}({\bf k},t)|^2 \rangle 
& = & 2(2\pi)^3\delta({\bf 0}) k_z^2V_{ii}({\bf k})
+2 \int^t_0dt'k_z^2 \delta \langle v_i^*({\bf k},t') v_i({\bf k},t)
\rangle b_0^2 \nonumber \\
& = & 4(2\pi)^3k_z^2J_1(k)b_0^2\delta^3({\bf 0})
-6(2\pi)^3 k_z^4b_0^4(\Delta t)_k^2J_1(k)\delta^3({\bf 0}),
\end{eqnarray}
%
where $(\Delta t)_k$ is the time scale of the eddy turnover and its expression 
will be given below. We assumed that 
the time integral should be estimated as 
$\int^t_0 dt' [\cdots] \sim (\Delta t)_k [\cdots] $ in the second line
of the right-hand side of the eq. (13) because the back-reaction works
only during the time scale of the eddy turnover.  
Here we assume Kolmogoroff spectrum for the inertial range
\footnote{The inertial range is defined by the scale which 
is smaller than the largest eddy ($\sim k_0^{-1}$) 
and larger than a small scale ($\sim R^{-3/4}k_0^{-1}$) 
under where the viscosity term is dominant. In this 
range, the transfer of the energy works from large eddy to small one  
without the dissipation of the energy. This leads a sort of 
`equilibrium state' with Kolmogoroff spectrum\cite{LL}(Kolmogoroff
Theory).} $k_0 < k < k_{\rm max} \sim R^{3/4}k_0$\cite{LL}, 
$k_0$ is the wave number of the largest eddy and $R$ is the Reynolds 
number. From the definition of the velocity dispersion  
%
\begin{eqnarray}
\langle v^2 \rangle =2 \int\frac{d^3k}{(2\pi)^3}J_1(k)\delta (0)
=: \int^{k_{\rm max}}_{k_0} dk I(k),
\end{eqnarray}
%
we obtain the relation 
%
\begin{eqnarray}
I(k) = \frac{1}{\pi^2} k^2 J_1(k) (\Delta t)_k^{-1} \simeq 
\frac{2}{3}v_0^2 \frac{k_0^{2/3}}{k^{5/3}},
\end{eqnarray}
%
where $v_0$ is the typical velocity($v_0 \sim {\sqrt {\langle
v^2\rangle}}$) and we used $\delta (0) 
\sim (\Delta t)_k^{-1}$. The expression of $(\Delta t)_k$ is 
given by the estimation of the order of the magnitude in the 
eq. (14), that is, $ (1/k(\Delta t)_k)^2 \sim k I(k)$.

Integrating the above equation (13) 
over ${\bf k}$, we obtain the modified 
Kulsrud and Anderson equation
%
\begin{eqnarray}
\partial_t \rho_M = 2 \gamma \rho_M -2 \zeta \rho_M^2,
\end{eqnarray}
%
where 
%
\begin{eqnarray}
\rho_M :=\frac{{\cal E}_M}{4\pi \rho}:=
\frac{1}{V}\int \frac{d^3k}{(2\pi)^3}\langle |{\bf b}({\bf k},t)|^2 \rangle 
\end{eqnarray}
%
%
\begin{eqnarray}
\gamma := 2 \int \frac{d^3k}{(2\pi)^3}k_z^2J_1(k)
\end{eqnarray}
%
and
%
\begin{eqnarray}
\zeta := 3 \int \frac{d^3k}{(2\pi)^3}k_z^4J_1(k)(\Delta t)_k^2.
\end{eqnarray}
%
In the above derivation, we used $\delta^3({\bf 0}) \sim V$, where 
$V$ is the typical volume of the system.   

Now we evaluate the coefficients $\gamma$ and $\zeta$. Results are
given by 
%
\begin{eqnarray}
\gamma \simeq \int^{k_{\rm max}}_{k_0} dk 
k^2I(k)(\Delta t)_k \simeq \int^{k_{\rm max}}_{k_0}dk [kI(k)]^{1/2}
\sim v_0k_0^{1/3}k_{\rm max}^{2/3} \sim R^{1/2}v_0k_0
\end{eqnarray}
%
and
%
\begin{eqnarray}
\zeta \simeq \int^{k_{\rm max}}_{k_0} dk k^4 I(k) (\Delta t)_k^3 \simeq  
\int^{k_{\rm max}}_{k_0} dk [kI(k)]^{-1/2} \sim 
\frac{k_{\rm max}^{4/3}}{v_0k_0^{1/3}}
\sim R\frac{k_0}{v_0},
\end{eqnarray}
%
respectively. Defining a dimensionless quantity $ \mu_M:=
\rho_M/v_0^2$, we can see that the eq.(16) becomes
%
\begin{eqnarray}
\partial_t \mu_M=2\gamma \mu_M-2 \zeta' \mu_M^2,
\end{eqnarray}
%
where $\zeta'=\zeta v_0^2 \sim R k_0 v_0$. The second term in the 
right-hand side of the eq. (22) comes from the effect of the 
back-reaction effect. One can see easily from the 
above equation that the back-reaction gives an opposite effect to the 
original kinetic term and 
make the energy of the magnetic field balance with the energy 
of the fluid. 

Although we know from the procedure used here that the eq. (22) holds only in 
a small time step as $\mu_M \ll 1$, we try to 
extrapolate. As a result we find the solution 
%
\begin{eqnarray}
\mu_M=\frac{\gamma}{\zeta'}\frac{1}{1-\Bigl(1-
\frac{1}{\mu_M(0)}\frac{\gamma}{\zeta'} \Bigr)e^{-2 \gamma t}}. 
\end{eqnarray}
%
One can see easily that the magnetic `energy' goes toward the terminal
value $\mu_M^*=\gamma /\zeta' \sim R^{-1/2}$ for a time scale 
$\sim \gamma^{-1}$.  
This value corresponds to the saturation value, which is estimated 
naively on the assumption that the drain by the magnetic field is 
comparable to the turbulent power\cite{Cen}\cite{KA}.

\section{Applications}

In this section we apply the eq. (22) to two examples which 
the magnetic field is amplified by the dynamo mechanism. First, we treat 
the time evolution of the magnetic field during  
the first order phase transition in the very early universe. 
We also consider briefly the
amplification of the magnetic field in interstellar mediums.   

\subsection{Electroweak Plasma}

There are attractive mechanisms of the generation of the 
primordial magnetic field in the course of cosmological  
phase transitions\cite{PT}. In these scenarios the strong magnetic 
field is expected to be amplified by MHD turbulence during the first 
order phase transition. The detail of the amplification 
has been discussed by using Kulsrud and 
Anderson equation in the ref. \cite{Olinto}.  

We reconsider the amplification of the magnetic field during 
the phase transition by using the modified Kulsrud and Anderson 
equation.  The time scale for the equipartition is 
$t_{\rm equi} \sim \gamma^{-1} \sim 
R^{-1/2}v_0^{-1}k_0^{-1}$. Since the Reynolds number is $R\sim 10^2$
\cite{Olinto}, we can see that it is the same order 
with the time scale of the phase transition. Thus, the 
magnetic field can be amplified enough and the final energy is 
given by  
%
\begin{eqnarray}
{\cal E}_M^* \sim R^{-1/2} {\cal E}_v \sim 0.1 \times {\cal E}_v,
\end{eqnarray}
%
where ${\cal E}_v$ is the energy of the plasma fluid. 

\subsection{Interstellar Mediums}

As we stated in Introduction, the kinetic dynamo theory breaks 
down in interstellar mediums\cite{KA}. For interstellar mediums, 
typical values of key quantities are $2\pi/k_0 \sim 100{\rm pc}$, $v_0 
\sim 10^6{\rm cm/s}$ and $R \sim v_0/k_0 \nu \sim 10^8$, where 
$\nu$ denotes the kinetic ion viscosity;
$\nu \sim 10^{18}{\rm cm}^2{\rm s}^{-1}$\cite{Cen}. 
Then the typical time scale 
is given by $t_{\rm ISM} \sim \gamma^{-1} \sim 10^2{\rm yr}$.  
Since the time scale of the mean field is $\sim 10^{10}{\rm yr}$\cite{KA}, 
we realize again the mean field theory is meaningless in the present 
perturbative approach. The final energy of the magnetic field is given by 
${\cal E}_M^* \sim 10^{-4} \times {\cal E}_v$. 

\section{Concluding Remark}

In this paper, we considered the lowest order back reaction to the 
kinetic dynamo theory and modified the equation for the 
energy of the magnetic field. As a result we 
obtained the successful time evolution of the energy of the 
magnetic field. 
That is, the terminal value of the magnetic energy obtained from 
the eq. (23) equals to the previous qualitative estimation  
of the saturation energy\cite{Cen}\cite{KA}. 
We also presented the expression depending on $k$ (eq. (13)), 
with which we can evaluate the evolution of the magnetic field 
for various scales. 
Since the present formalism is general, our equation is useful 
for other situations, for example, the fireball model for 
$\gamma$-ray bursts\cite{Gamma}.  

Finally, we should comment on our assumption for the initial 
condition (eq. (12)) 
and the extrapolation of the eq. (22). We choose the initial condition 
in order to obtain the simple result 
like the eq. (22). Although this assumption holds approximately in 
some cases, it may not be correct in general cases. 
We should also note that we considered only the effect of lowest order
back-reaction. Properly 
speaking, if one wishes to analyse the vicinity of the equipartition, 
one must take account of effects of higher order back-reaction. 
The study near the equipartition might become clear by using 
something like the renormalization group approach. 
The study for more general initial 
condition and with higher order back reaction should be done in the  
future. At the same time, the spatial structure as the typical 
coherent length of the magnetic field also should be discussed.

\section*{Acknowledgements}
We would like to thank Katsuhiko Sato for his continuous 
encouragement and Masahiro Morikawa for his comment. TS  
is grateful to Gary Gibbons and DAMTP relativity group for their 
hospitality. We also thanks T. Uesugi for a careful reading of the 
manuscript of thispaper. This work was partially supported by the Japanese 
Grant-in-Aid for Scientific Research on Priority Areas (No. 10147105) 
of the Ministry of Education, Science, Sports, and Culture 
and Grant-in-Aid for Scientific Research from the Ministry of E
ducation, Science, Sports, and Culture, No. 08740170 (RN).

\end{document}